\documentstyle[12pt,epsfig]{article}

\newlength{\dinwidth}
\newlength{\dinmargin}
\def\lapproxeq{\lower .7ex\hbox{$\;\stackrel{\textstyle
<}{\sim}\;$}}
\def\gapproxeq{\lower .7ex\hbox{$\;\stackrel{\textstyle
>}{\sim}\;$}}

\def\be{\begin{equation}}
\def\ee{\end{equation}}
\def\bea{\begin{eqnarray}}
\def\eea{\end{eqnarray}}

\def\funp{{I\!\!P}}
\def\gtrsim{ \;\raisebox{-.7ex}{$\stackrel{\textstyle
>}{\sim}$}\; }
\def\lesim{ \;\raisebox{-.7ex}{$\stackrel{\textstyle
<}{\sim}$}\; }

\def\ra{ \rightarrow }

\begin{document}
\titlepage
\begin{flushright}
IPPP/02/18 \\
DCPT/02/36 \\
22 May 2002 \\
\end{flushright}

\vspace*{2cm}

\begin{center}
{\Large \bf Physics with tagged forward protons at the LHC}

\vspace*{1cm}
V.A. Khoze$^{a,b}$, A.D. Martin$^a$,  and M.G. Ryskin$^{a,b}$ \\

\vspace*{0.5cm} $^a$ Department of Physics and Institute for
Particle Physics Phenomenology,\\ University of
Durham, Durham, DH1 3LE \\
$^b$ Petersburg Nuclear Physics Institute, Gatchina,
St.~Petersburg, 188300, Russia
\end{center}

\vspace*{1cm}

\begin{abstract}
We emphasize the importance of tagging the outgoing forward
protons to sharpen the predictions for New Physics at the LHC
(such as the diffractive production of a light Higgs boson). The
rescattering effects lead to a rich distinctive structure of the
cross section as a function of the transverse momenta of the
protons. We show that a study of the correlations between the
proton transverse momenta for double-diffractive production of
central dijets will provide a detailed check of the whole
diffractive formalism. Adopting a perturbative two-gluon structure
of the Pomeron, we emphasize that $2^{++}$ quarkonium production,
via Pomeron-Pomeron fusion, is strongly suppressed. This offers a
favourable production mechanism for non-$q\bar{q}$ states, such as
glueballs.
\end{abstract}

\newpage
\section{Introduction}

Exclusive double-diffractive-like processes of the type
\begin{equation}
\label{eq:a1}
 pp \ra p + M + p
\end{equation}
can significantly increase the physics potential of high energy
proton colliders. Here $M$ represents a system of invariant mass
$M$, and the + signs denote the presence of rapidity gaps which
separate the system $M$ from the protons. Such processes allow, on
the one hand, novel studies of QCD and of the diffractive
amplitude at very high energies, while, on the other hand, allow
an exceptionally clean experimental environment to identify New
Physics signals (such as the Higgs boson, SUSY particles, etc.,
see \cite{KMR} and ref. therein). Moreover tagging two forward
protons offers an attractive extension of the proton collider
physics programme to studies of high-energy $\gamma\gamma$
collision physics; see, for example, [1--3].

In such events we produce a colour-singlet state $M$ which is
practically free from soft secondary particles. Moreover, if
forward going protons are tagged we can reconstruct the `missing'
mass $M$ with good resolution, and so have an ideal means to
search for new resonances and to study threshold behaviour
phenomena. We have to pay a price for ensuring such a clean
diffractive signal. In particular, the diffractive event rate is
suppressed by the small probability, $\hat{S}^2$, that the
rapidity gaps survive soft rescattering effects between the
interacting hadrons, which can generate secondary particles which
populate the gaps [4--12].

In general, we may write the survival factor $\hat{S}^2$ in a
multi-channel eikonal framework in the form
\begin{equation}
\label{eq:a2}
 \hat{S}^2 \; = \; \frac{\int \sum_i \left |{\cal M}_i (s, b_t^2)
 \right |^2 \: \exp \left (- \Omega_i (s, b_t^2) \right ) d^2 b_t}{\int
 \sum_i \left | {\cal M}_i (s, b_t^2) \right |^2 d^2 b_t}
\end{equation}
where the incoming proton is decomposed into diffractive
eigenstates, each with its own opacity\footnote{Really we deal
with a matrix $\Omega_{jj^\prime}^{ii^\prime}$, where the indices
refer to the eigenstates of the two incoming and two outgoing
hadrons \cite{KKMR}.} $\Omega_i$.  The amplitudes ${\cal M}_i
(s,b_t^2)$ of the process of interest may be different in the
different diffractive eigenstates.  They are expressed in impact
parameter $b_t$ space at centre-of-mass energy $\sqrt{s}$.  It is
important to recall that the suppression factor $\hat{S}^2$ is not
universal, but depends on the particular hard subprocess, as well
as on the kinematical configurations of the parent reaction, see,
for example, \cite{KKMR}.

Double-diffractive Higgs production,
\begin{equation}
\label{eq:a3}
 pp \ra p + H + p,
\end{equation}
at the LHC, is a good example of illustrating the pros and cons of
such exclusive processes. Let us assume a Higgs boson of mass $M_H
= 120\:{\rm GeV}$ and consider detection in the $b\bar{b}$ decay
channel. The disadvantage is that to ensure the survival of the
rapidity gaps in (\ref{eq:a3}), the predicted cross section is
low, $\sigma\simeq 2\:{\rm fb}$, corresponding to a survival
factor $\hat{S}^2=0.02$. The advantage is that, {\em by tagging
the outgoing protons}, the signal-to-background ratio\footnote{The
ratio quoted in (\ref{eq:a4A}) corresponds to a cut on the $b$ jet
transverse momenta of $p_t(b)>0.4M_H$.} is extremely favourable
relative to other Higgs signals \cite{KMRmm,VAK01},
\begin{equation}
\label{eq:a4A} \frac{{\rm signal\ }(H\ra b{\bar b})}{b{\bar b}\
{\rm QCD\ background}}\gtrsim 4,
\end{equation}
if the missing mass resolution $\Delta M$ obtained by the proton
taggers is $1\:{\rm GeV}$. Indeed, with dedicated forward proton
spectrometers at the LHC, the process $pp\ra p + H + p$ may even
be the light Higgs discovery channel.

For completeness, we note that for Higgs production via
photon-photon fusion, the survival factor is much larger,
$\hat{S}^2\simeq 0.86$ \cite{KMR,KMRb}, and the corresponding
cross section $\sigma_{\gamma\gamma}(H\ra b\bar{b})\simeq0.1\:{\rm
fb}$. In this case the signal-to-background ratio is
\begin{equation}
\label{eq:a4B} \frac{{\rm signal\ }(\gamma\gamma\ra H\ra
b\bar{b})}{b\bar{b}\ {\rm QED\ background}}\ \simeq\ \frac{7\:{\rm
GeV}}{\Delta M_{b\bar{b}}}\ .
\end{equation}

There are other physics reasons why it is desirable to tag the
recoil protons in double diffractive processes. First, it offers a
valuable experimental probe of the opacities $\Omega_i(s,b_t^2)$
of the proton. The relevant Feynman diagrams for process
(\ref{eq:a1}) are shown in Fig.~1. There is appreciable
interference between the amplitudes without and with the soft
rescattering corrections, which are shown in Figs. 1(a) and 1(b)
respectively. We will show that the interference depends on the
transverse momenta ${\vec p}_{1t}, {\vec p}_{2t}$ of the recoil
protons and on the azimuthal angle $\phi$ between these momenta.
This dependence can be used to probe\footnote{Another possibility,
to probe the opacity of the proton, is to study the process with
rapidity gaps mediated by photon exchange \cite{KMRb}. By varying
the momentum transfer of the photon we sample different impact
parameters $b_t$ and hence scan the opacity $\Omega(s,b_t^2)$.}
the different soft rescattering models and the behaviour of the
opacities $\Omega_i(s,b_t^2)$.

\begin{figure}
\begin{center}
\epsfig{figure=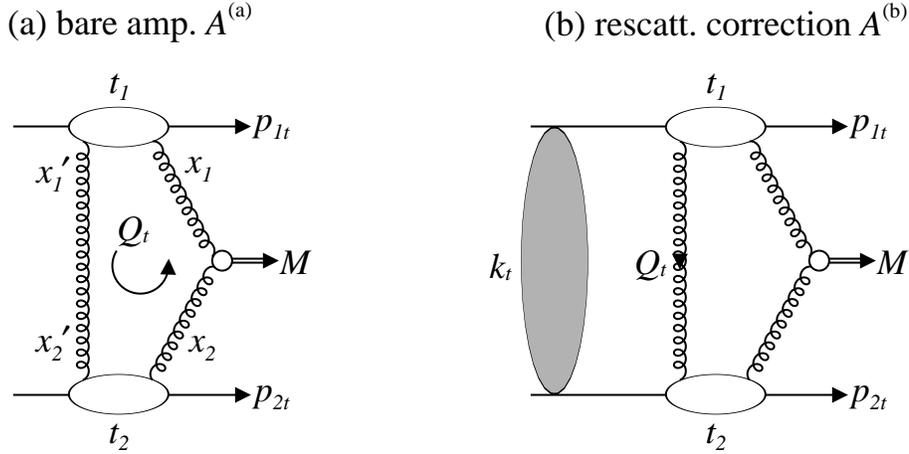,height=2.5in} \caption{The bare amplitude
$A^{(a)}$ and the rescattering correction $A^{(b)}$ for the
double-diffractive process $pp\ra p+M+p$.}
 \label{Fig1}
\end{center}
\end{figure}

Secondly, we need to understand, and if possible to predict, the
${\vec p}_{it}$ behaviour of the diffractive cross sections in
order to plan experiments and to evaluate the acceptance and
efficiency of the leading proton detectors at the LHC.

The content of the paper is as follows. In Section~2 we recall how
the bare perturbative amplitude $A^{(a)}$ of Fig.~1(a) may be
calculated, and then in Section~3 we illustrate the effect of
rescattering corrections in terms of a simple model. Throughout
the paper, it is safe to neglect the rescattering of the centrally
produced system $M$ on either proton. This system (for example, a
Higgs boson, high $E_T$ dijet, etc.) is massive and has small
size, so, due to colour transparency \cite{KT}, the cross section
$\sigma(Mp)$ is very small. Rescattering is computed realistically
in Section~4 and the predictions for a general double-diffractive
process, $pp\ra p+M+p$, are presented in Section~5. A detour is
made in Section~6 to discuss double-diffractive light meson
production, for which data already exist. Of course, this is
beyond the region of validity of a perturbative QCD approach, but,
surprisingly, the perturbative predictions agree qualitatively
with interesting features of these data. Our conclusions are
presented in Section~7.

\section{The bare amplitude}

The amplitude $A^{(a)}$ of Fig.~1(a), describing the high energy
double-diffractive production of a heavy system $M$, can be
expressed in terms of the generalised (skewed) unintegrated gluon
densities $f_g(x,x^\prime,t,Q_t,\mu)$. Here $\mu\simeq M/2$ is the
scale of the hard $gg\ra M$ subprocess, and $t$ is the transverse
momentum squared transferred through the `hard' QCD Pomeron (that
is the two-gluon system). Essentially the gluon distribution $f_g$
opens up and describes the internal structure of the `hard' QCD
Pomeron, whose exchange mediates the diffractive process
(\ref{eq:a1}).

For the exclusive reaction (\ref{eq:a1}) the bare amplitude of
Fig.~1(a) is, to single log accuracy, given by \cite{KMRhiggs}
\begin{equation}
\label{eq:a4} A^{(a)}=\frac{1}{N_c^2-1}\int\frac{d^2Q_t}{Q_t^4}
f_g\left(x_1,\dots Q_t,\mu\right)f_g\left(x_2,\dots
Q_t,\mu\right){\cal M}
\end{equation}
where ${\mathcal M}$ is the matrix element of the hard $gg\ra M$
subprocess. For example, the cross section for the $gg\ra gg$
subprocess, relevant to high $E_T$ dijet production
\cite{KMR,KMRdijet,BC}, is

\begin{equation}
\label{eq:a5} \frac{d\hat{\sigma}}{dt}= |{\cal M}|^2 = \frac{9}{4}
\frac{\pi\alpha_s^2}{E_T^4}.
\end{equation}

For small $|x_i-x_i^\prime|$, which is appropriate for high energy
double diffraction, and $t=0$, the skewed unintegrated density
$f_g$ can be calculated from knowledge of the conventional
integrated gluon \cite{MR,SGMR}. The precise form of the $t$
dependence of $f_g$ is not well known. Recall, however, that
$f_g\left(\dots Q_t,\mu\right)$ contains a Sudakov-like factor
$T(Q_t,\mu)$ which reflects the chance that a gluon with
transverse momentum $Q_t$ remains untouched in the evolution up to
the hard scale $\mu$---a necessary condition for the survival of
the rapidity gap, see, for example, \cite{KMRhiggs,KMRdijet,KMRH}.
It is this $T$ factor which provides the infrared stability of the
$Q_t$ integral of (\ref{eq:a4})\footnote{Moreover, the effective
anomalous dimension of the gluon distribution additionally
suppresses the contribution from the low $Q_t$ domain
\cite{KMRH}.}. For example, for the production of a system of mass
$M\gtrsim 100\:{\rm GeV}$ at the LHC, the saddle point of the
$Q_t^2$ integration occurs at $Q_t^2\gtrsim 3$--$4\:{\rm GeV}^2$.
On the other hand, the transferred momenta satisfy $|t_i|\lesim
0.5\:{\rm GeV}^2$, which are small in comparison with $Q_t^2$.
Therefore it is natural to assume the factorization
\begin{equation}
\label{eq:a6}f_g\left(x,x^\prime,t,Q_t,\mu\right) =
\beta(t)f\left(x,x^\prime,t=0,Q_t,\mu\right),
\end{equation}
where the whole $t$ dependence is given by the effective
form-factor $\beta(t)$ of the QCD Pomeron-proton vertex. In other
words, we separate the dependence of the $pp\ra p+M+p\,$ cross
section on the transverse variables ($t_1,t_2$ or ${\vec
p}_{1t},{\vec p}_{2t}$) from the dependence on the `longitudinal'
variables (the initial $pp$ energy $\sqrt{s}$, the mass $M$ and
rapidity $y$ of the system $M$). That is the bare amplitude is
given by
\begin{equation}
\label{eq:a7}A^{(a)}\left({\vec p}_{1t},{\vec p}_{2t}\right) =
\beta(t_1)\beta(t_2)A_M
\end{equation}
where $t_i \simeq -p_{it}^2$, and where $A_M\equiv
A^{(a)}\left(p_{1t}=p_{2t}=0\right)$ may be calculated from
(\ref{eq:a4}).

\section{Absorption correction: a first look}

To calculate the absorptive or soft rescattering amplitude
$A^{(b)}$ of Fig.~1(b), it is convenient to use the momentum
representation. We may neglect the spin-flip component in the
proton-Pomeron vertex\footnote{This component is expected to be
small and consistent with zero.  If we note the similarity between
the photon and Pomeron vertices then the magnitude of the
isosinglet spin-flip amplitude is proportional to $\left
|\frac{1}{2} (\mu_p^a + \mu_n^a) \right | \lapproxeq 0.06$, where
the anomalous magnetic moments $\mu^a$ of the neutron and proton
cancel each other almost exactly.}. We perform the detailed
calculation in Section 4, but, first, we estimate the qualitative
features of the rescattering effect by assuming, in this Section,
that the amplitude for elastic proton-proton scattering, at energy
$\sqrt{s}$ and momentum transfer $k_t$, has the simplified form
\begin{equation}
\label{eq:a8}
 A_{pp} (s, k_t^2) \; = \; A_0 (s) \: \exp (-B k_t^2/2).
\end{equation}
From the optical theorem we have ${\rm Im} A_0 (s) = s
\sigma_{pp}^{\rm tot} (s)$, and for the small contribution of the
real part it is sufficient to use ${\rm Re} A_0/{\rm Im} A_0
\simeq 0.12$ in the energy regime of interest.  $B$ is the slope
of the elastic $pp$ differential cross section, $d \sigma_{pp}/dt
\propto \exp (Bt)$.

Using the above elastic $pp$ amplitude we may write the
rescattering contribution, Fig.~1(b), to the $pp \ra p + M + p$
amplitude as
\begin{equation}
\label{eq:a9}A^{(b)} =
i\int\frac{d^2k_t}{8\pi^2}\beta(t_1)\beta(t_2)A_M\frac{A_0}{s}e^{-Bk_t^2/2}
\end{equation}
where now $t_1\simeq-({\vec k}_t - {\vec p}_{1t})^2$ and
$t_2\simeq-({\vec k}_t + {\vec p}_{2t})^2$. If we take an
exponential form for the QCD Pomeron vertices,
\begin{equation}
\label{eq:a10}\beta(t) = e^{bt/2},
\end{equation}
then the integral in (\ref{eq:a9}) can be evaluated, to give
\begin{equation}
\label{eq:a11}A^{(b)}({\vec p}_{1t},{\vec p}_{2t}) =
\frac{iA_0}{4\pi s(B+2b)} \exp\left(\frac{b^2|{\vec p}_{1t}-{\vec
p}_{2t}|^2}{2(B+2b)}\right)A^{(a)}({\vec p}_{1t},{\vec p}_{2t}).
\end{equation}

To gain insight it is useful to compute the numerical value of
$A^{(b)}$ at the LHC energy using reasonable values of the
parameters. We take $b=4\:{\rm GeV}^{-2}$, $B=20\:{\rm GeV}^{-2}$,
$\sigma_{pp}^{\rm tot}=100\:{\rm mb}$ and, for the moment, neglect
the real part, ${\rm Re}A_0$, of the $pp$ elastic amplitude. We
obtain
\begin{equation}
\label{eq:a12}A^{(b)}=-0.73\exp \left(C|{\vec p}_{1t}-{\vec
p}_{2t}|^2\right)A^{(a)},
\end{equation}
where $C=0.29\:{\rm GeV}^{-2}$. Thus in the back-to back
configuration with ${\vec p}_{1t} \sim -{\vec p}_{2t} \sim
0.5\:{\rm GeV}$, the absorptive correction $A^{(b)}$ completely
cancels the bare amplitude $A^{(a)}$, and we predict a deep
diffractive dip. Moreover we see that the position of the dip
depends on the azimuthal angle $\phi$ between the transverse
momenta ${\vec p}_{1t}$ and ${\vec p}_{2t}$ of the tagged protons.
For $\phi = 180^\circ$ the momentum transfer occurs mainly through
the elastic amplitude $A_{pp}$, with $|t_1|$ and $|t_2|$ minimized
simultaneously, and hence the amplitude $A^{(b)}$ becomes larger.

\section{Detailed treatment of rescattering corrections}

To make a realistic calculation of the rescattering corrections we
must improve the description of $pp$ soft interaction. We use the
model of Ref.~\cite{KMRsoft}. It embodies (i)~pion-loop insertions
to the Pomeron trajectory, (ii)~two-channel eikonal description of
proton-proton rescattering and (iii) high mass diffractive
dissociation.  The parameters of the model were tuned to describe
all the main features of the soft $pp$ data throughout the
CERN-ISR to the Tevatron energy interval.  In terms of the two
channel eikonal the incoming proton is described by two
diffractive eigenstates $|\phi_i \rangle$, each with its own
absorptive cross section.

The eigenstates were taken to have the same profile in impact
parameter space, and absorptive cross sections
\begin{equation}
\label{eq:a13}
 \sigma_i \; = \; a_i \sigma_0 \quad\quad {\rm with} \quad\quad
 a_i \; = \; 1 \pm \gamma,
\end{equation}
where $\gamma = 0.4$ and $\sigma_0$ is defined in
Ref.~\cite{KMRsoft}. That is the two channel opacity is
\begin{equation}
\label{eq:a14}
 \Omega_{jj^\prime}^{ii^\prime} \; = \; \delta_{ii^\prime}
 \delta_{jj^\prime} a_i a_j \Omega.
\end{equation}
The impact parameter representation of the elastic amplitude is
thus
\begin{equation}
\label{eq:a15} {\rm Im} \: \tilde{A}_{pp} (s,b_t) \; = \; s\left(
1 \:
 - \: \frac{1}{4} \left [e^{- (1 + \gamma)^2 \Omega/2} \: + \:
 2e^{- (1 - \gamma^2) \Omega/2} \: + \: e^{- (1 - \gamma)^2
 \Omega/2}\right ] \right ).
\end{equation}
When we allow for the extra $(1 \pm \gamma)^2$ factors, which
reflect the different Pomeron couplings to the two diffractive
eigenstates in the $pp \rightarrow p + M + p$ production
amplitude, that is in the right-hand part of Fig.~1(b), we obtain
the effective amplitude of $pp$ rescattering,
\begin{equation}
\label{eq:a17} {\rm Im}\:\tilde{A}_{pp} (s, b_t) \; = \; s \left (
1 \: - \: \frac{1}{4}
 \left [ (1 + \gamma)^2 e^{- (1 + \gamma)^2 \Omega/2} \: + \: 2(1-\gamma^2)e^{-(1 -
 \gamma^2) \Omega/2}\: + \: (1 - \gamma)^2 e^{- (1 - \gamma)^2
 \Omega/2}\right ] \right ).
\end{equation}
The optical density $\Omega (s, b_t^2)$ was given in
Ref.~\cite{KMRsoft} for Tevatron $(\sqrt{s} = 2~{\rm TeV})$ and
LHC $(\sqrt{s} = 14~{\rm TeV})$ energies.

As before, we work in momentum space, and replace (\ref{eq:a8}) by
\begin{equation}
\label{eq:a18}
 A_{pp} (s, k_t^2) \; = \; \frac{1}{2 \pi} \int d^2 b_t \,
 4\pi \,  \tilde{A}_{pp} (s, b_t) \: e^{i \vec{k}_t \cdot \vec{b}_t}.
\end{equation}
In this way we obtain a more realistic evaluation of the
rescattering amplitude $A^{(b)}$ of Fig.~1(b). However, from the
na\"{\i}ve evaluation of Section~3, we anticipate that there will
still be a strong cancellation between the bare amplitude
$A^{(a)}$ and the absorptive correction $A^{(b)}$, originating
from the imaginary part of the elastic rescattering $A_{pp}$.

First, we must introduce the real part of $A_{pp}(s,k_t^2)$.
However, it is not justified to use a constant ratio ${\rm
Re}\:A_{pp}\: / \: {\rm Im}\:A_{pp} \: \simeq \: 0.12$ for $k_t$
away from zero. To account for the $k_t$ dependence we use the
dispersion relation result, recalling that the total $pp$ cross
section increases logarithmically with energy. Then the
even-signature amplitude satisfies
\begin{equation}
\label{eq:a19}{\rm Re}\:\tilde{A}_{pp}(s,b_t)\ = \
\frac{\pi}{2}\:s\:\frac{\partial\left({\rm
Im}\:\tilde{A}_{pp}(s,b_t)/s\right)}{\partial\ln s}\:.
\end{equation}
To convert this relation from the impact parameter space amplitude
$\tilde{A}_{pp}(s,b_t)$ to the momentum space amplitude
$A_{pp}(s,k_t^2)$ we use (\ref{eq:a18}).

Second, we should specify the form of the QCD Pomeron-proton
vertex, $\beta(t)$. The most consistent choice is to take the
dipole form used in Ref.~\cite{KMRsoft}
\begin{equation}
\label{eq:a20}\beta(t)=\frac{1}{(1-t/a_1)}\frac{1}{(1-t/a_2)}\ .
\end{equation}
For comparison we also evaluated the $\: pp\ra p+M+p\:$ cross
section using the alternative form
\begin{equation}
\label{eq:a21}\beta(t) = e^{bt/2}
\end{equation}
with $b=4\:{\rm GeV}^{-2}$, which is consistent with the $\:\gamma
p\ra J/\psi + p\:$ HERA data\footnote{The latest HERA data for
$J/\psi$ elastic photoproduction prefer $b=4.5\:{\rm GeV}^{-2}$ .
However, for a heavier system, the smaller slope $b=4\:{\rm
GeV}^{-2}$ looks more reasonable.} \cite{JPSI, LEVY}.

\section{Predictions for tagged protons at the LHC}

We are now in a position to predict the transverse momentum
dependence of the outgoing protons in the double-diffractive
production of a heavy system of mass $M$, that is in the process
$pp\ra p + M + p$. We show the results in the form
\begin{equation}
\label{eq:a22}M^2 \frac{\partial^4{\cal L}}{\partial y\partial M^2
\partial^2p_{1t}\partial^2p_{2t}} = M^2\frac{\partial^2{\cal L}}
{\partial y \partial M^2}F(\vec{p}_{1t},\vec{p}_{2t})
\end{equation}
where $M^2 d^2{\cal L}/dydM^2$ is the Pomeron-Pomeron luminosity
given in Ref.~\cite{KMR} and where the factor $F$ contains the
explicit $\vec{p}_{1t}$ and $\vec{p}_{2t}$ dependence. The
luminosity, given in \cite{KMR}, was integrated over
$d^2p_{1t}d^2p_{2t}$ with the assumption that the bare amplitude
had an exponential $t$ behaviour,
\begin{equation}
\label{eq:a23} A^{(a)}\propto \exp\left(b_0(t_1+t_2)/2\right)
\end{equation}
with $b_0 = 4\:{\rm GeV}^{-2}$. In addition, the soft survival
probability $\langle\,S^2\,\rangle$ was averaged over the
available $t_1,t_2$ domain. Here we unfold the luminosity to
expose the $\vec{p}_{1t}$ and $\vec{p}_{2t}$ dependence. In order
to be able to use the published $M^2\partial^2{\cal L}/\partial
y\partial M^2$ luminosity, we therefore compute
\begin{equation}
\label{eq:a24} F(\vec{p}_{1t},\vec{p}_{2t}) \; = \;
\frac{\beta^2(t_1)\beta^2(t_2)} {\langle S^2\rangle\pi^2/b_0^2}
\;\frac{\partial^2S^2(\vec{p}_{1t},\vec{p}_{2t})}
{\partial^2p_{1t}\partial^2p_{2t}}\:.
\end{equation}
From the product of $F$, computed in this way, and the luminosity
given in Ref.~\cite{KMR}, we obtain the luminosity as a function
of $\vec{p}_{1t}$ and $\vec{p}_{2t}$, as well as of $y$ and $M$.
This resultant luminosity, (\ref{eq:a22}), needs only be
multiplied by the appropriate hard subprocess $gg^{PP}\ra M$ cross
section\footnote{The notation $gg^{PP}$ is to indicate that the
hard gluons, which interact to form the system $M$, originate
within overall (colourless) hard Pomeron exchanges.} to obtain the
differential cross section for any $pp\ra p+M+p$ diffractive
process. Various hard subprocess cross sections were listed, and
discussed, in Ref.~\cite{KMR}.

\begin{figure}
\begin{center}
\epsfig{figure=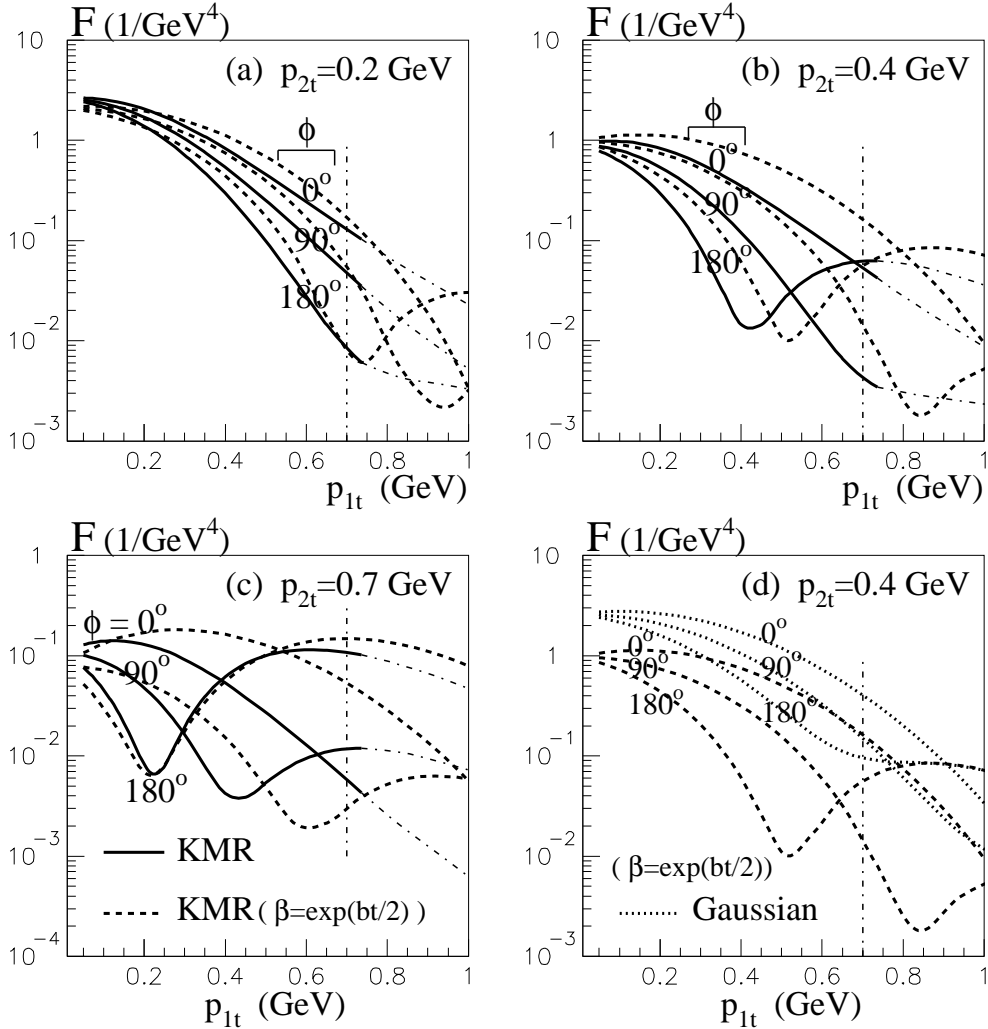,height=6in} \caption{The factor
$F(\vec{p}_{1t},\vec{p}_{2t})$ of (\ref{eq:a22}) and
(\ref{eq:a24}), which specifies the forward going proton
transverse momentum dependence of the $pp\ra p+M+p$ cross section,
for typical values of $p_{1t},p_{2t}$ and the azimuthal angle
$\phi$. The first three plots correspond to
$p_{2t}=0.2,0.4,0.7\:{\rm GeV}$ respectively, and show the results
obtained using the KMR two-channel eikonal model of
Ref.~\cite{KMRsoft} to calculate the soft rescattering, as
described in Section~4. The dashed curves show the sensitivity to
the form of the QCD Pomeron-proton vertex $\beta(t)$, by replacing
the dipole form (\ref{eq:a20}) by the exponential form
(\ref{eq:a21}). The dotted curves in Fig.(d) correspond to the use
of a na\"{\i}ve single-channel eikonal model (with $A^{(b)}$
computed from (\ref{eq:a8}) and (\ref{eq:a9})) as compared to that
obtained with the `realistic' two-channel eikonal model of
Ref.~\cite{KMRsoft}; in both cases the exponential form factor was
used, so the dashed curves are the same in plots (d) and (b).}
 \label{Fig2}
\end{center}
\end{figure}

The factor $F$ is plotted in Figs.~2(a,b,c) as a function of
$p_{1t}$ for three values of $p_{2t}=0.2, 0.4$ and $0.7\:{\rm
GeV}$ respectively. In each case the factor is shown for $\phi =
0^\circ, 90^\circ$ and $180^\circ$, where $\phi$ is the azimuthal
angle between $\vec{p}_{1t}$ and $\vec{p}_{2t}$. The continuous
and dashed curves are obtained using (\ref{eq:a20}) and
(\ref{eq:a21}), respectively, for the QCD Pomeron-proton vertex.
Recall that the model of Ref.~\cite{KMRsoft} was fitted to `soft'
diffractive $pp$ data in the region $|t|\leq 0.5\:{\rm GeV}^2$,
and so, strictly speaking, $\beta (t)$ of (\ref{eq:a20}) should
only be applied for $p_t\lesim 0.7\:{\rm GeV}$. However, we hope
we can reliably evaluate rescattering corrections up to $p_t\simeq
1\:{\rm GeV}$, as the typical values of $k_t$ (the momentum
transferred through the elastic amplitude $A_{pp}$ in Fig.~1(b)),
which are controlled mainly by the elastic slope $B/2$, are much
less than $1\:{\rm GeV}$. As discussed before, the absorptive
corrections are stronger in the back-to-back configuration;
already for $p_t < 0.7\:{\rm GeV}$ the $\phi=180^\circ$ curves
reveal a rich dip structure.

In the final plot, Fig.~2(d), we compare the prediction for $F$
obtained using the elastic amplitude determined in the two-channel
eikonal model of Ref.~\cite{KMRsoft}, with a na\"{\i}ve estimate
based on a simple one-channel approach where the elastic $pp$
amplitude is given by the Gaussian formula of (\ref{eq:a8}), that
is the amplitude is described by single Pomeron exchange. However,
we keep the parameters found in \cite{KMRsoft}, that is
$\sigma_{pp}^{\rm tot} = 102\:{\rm mb}$, $B=20.7\:{\rm GeV}^{-2}$
and ${\rm Re}A_{pp}/{\rm Im}A_{pp}= 0.12$ at $k_t = 0$. In both
cases we use (\ref{eq:a21}) for $\beta(t)$. The large difference
between the realistic and na\"{\i}ve predictions demonstrates
their sensitivity to the model used for soft rescattering.

In Fig.~3 we show the azimuthal dependence of the `soft' survival
factor
\begin{equation}
\label{eq:a25} S^2(\vec{p}_{1t},\vec{p}_{2t}) =
\frac{|A^{(a)}+A^{(b)}|^2}{|A^{(a)}|^2}\:,
\end{equation}
as a function of the azimuthal angle $\phi$, for different choices
of $p_{1t}$ and $p_{2t}$. The rich structure of $S^2$ is apparent,
which feeds through into the double-diffractive cross section. As
anticipated, we observe a flatter behaviour in $\phi$ for small
$p_{1t}$ and $p_{2t}$, while for larger $p_t \sim 0.7\:{\rm GeV}$
a diffractive dip already occurs for $\phi \sim 90^\circ$.

\begin{figure}
\begin{center}
\epsfig{figure=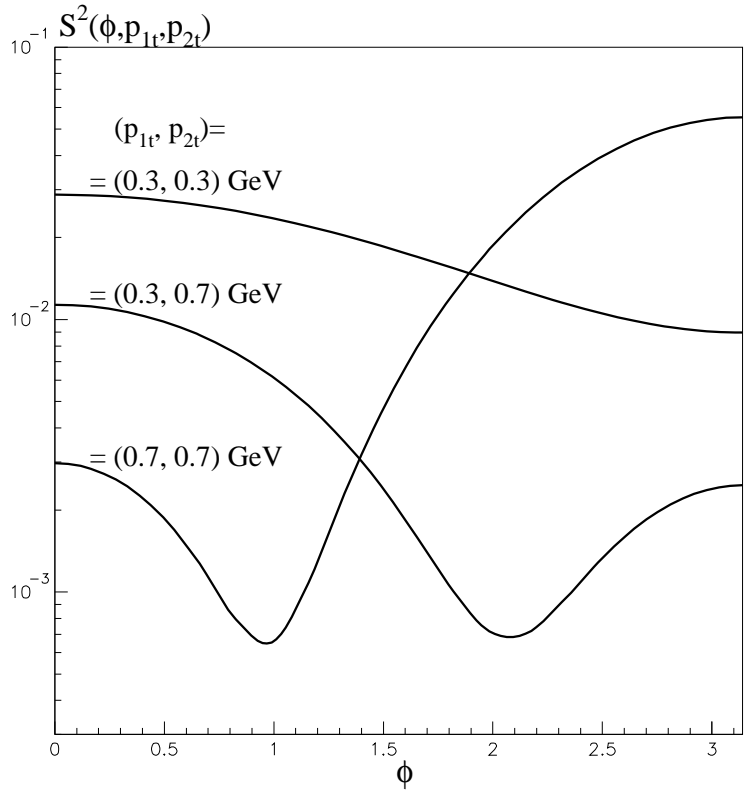,height=4.5in} \caption{The dependence of
the survival probability, $S^2$, of the rapidity gaps on the
azimuthal angle $\phi$ between the transverse momenta
$\vec{p}_{it}$ of the forward going protons in the process $pp\ra
p+M+p$, for typical values of $p_{1t}$ and $p_{2t}$.}
 \label{Fig3}
\end{center}
\end{figure}

\section{Application to double diffractive meson production}

An interesting $\phi$ behaviour has been observed by the WA102
collaboration at CERN \cite{WA102} at lower energies ($\sqrt s
\sim 30\:{\rm GeV}$) in fixed target central double-diffractive
meson production,
\begin{equation}
\label{eq:a26} pp\ra p+X+p,
\end{equation}
where a partial wave analysis of the $X$ channel allows the
identification of a wide range of meson resonances.

It has been emphasized in Ref.~\cite{CLOSE} that within the Regge
framework the Reggeon-Reggeon$\:\ra\:$meson vertex, $V(RR\ra X)$,
embodied in the amplitude $A^{(a)}$, may contain an azimuthal
dependence like
\begin{equation}
\label{eq:a27} |V|^2 = 1+a\cos\phi+b\cos^2\phi.
\end{equation}
In fact, with an appropriate choice of parameters, a
phenomenological description of the $\phi$ dependences observed in
the meson-production data can be achieved without any rescattering
corrections \cite{CLOSE}.

For a hard subprocess at scale $\mu=M/2$ with $Q_t^2\gg t_i$ we
have no such dependence in the vertex $V(\funp\funp\ra M)$; the
coefficients in (\ref{eq:a27}) satisfy $a,b\lesim|t_i|/4Q_t^2$. On
the other hand, at least part of the azimuthal effects observed in
the meson data \cite{WA102} may originate in the rescattering
corrections discussed in the present paper.

To study this further, we adopt the perturbative QCD viewpoint,
which offers a dynamical basis in which to understand the
structure of process (\ref{eq:a26}). Of course, it is questionable
to use perturbative QCD to describe the production of rather light
mesons, via (\ref{eq:a26}), where we have no hard scale. On the
other hand it is natural to expect a smooth matching between the
`soft' and perturbative regimes. In this way we may obtain a
qualitative interpretation of observed features the data. As we
shall see below, this indeed turns out to be the case.

Recall that, in general, for forward going protons ($p_{it} \ll
Q_t$), two `hard' QCD Pomerons can produce only a P-even state
with the longitudinal projection of its spin $J_z=0\,$
\cite{KMRmm,KMRL}. Also note that, as the QCD Pomerons are built
from gluons, the underlying fusion subprocess $gg^{PP}\ra X$ may
provide a favourable environment for the production of exotic
meson states containing gluons (such as glueballs, hybrids, etc.);
the cross section of the $gg^{PP}\ra q\bar{q}$ subprocess is much
smaller than that of $gg^{PP}\ra gg$, especially in the $J_z=0$
channel. Next for $J_z=0$ the vertex $gg^{PP}\ra X(2^{++})$ is
strongly suppressed if the $2^{++}$ state is a normal
non-relativistic $q\bar{q}$ meson\footnote{The origin of this
result can actually be traced to the absence of the $\gamma\gamma$
decay mode of $2^{++}$ positronium in the $J_z=0$ state
\cite{ALEK}, and then to the absence of the $gg$ decay of a
$J_z=0\;$ $q\bar{q}$ system \cite{JGG}.}. This is the result of
gauge invariance. Indeed, for $X$ made from a non-relativistic
$q\bar{q}$ pair, the fusion process $gg^{PP}\ra(q\bar{q})$ looks
like a single local vertex. The distance between the two gluon
vertices is of the order of the inverse constituent quark mass
($1/m_q$), which is much smaller than the size of the ($q\bar{q}$)
bound state. Now, the structure of the local $gg^{PP}\ra
X(2^{++})$ interaction is fixed by gluon gauge invariance. Then
the requirement that the polarization tensor $T_{\mu\nu}$ of the
$2^{++}$ meson satisfies $T_{\mu\mu}=0\,$ means that the vertex
vanishes for $J_z=0$ \cite{TEN}. The consequence is that the
forward double-diffractive production of normal quarkonium
$q\bar{q} (2^{++})$ states should be suppressed \cite{KMRmm,TEN},
and so Pomeron-Pomeron fusion produces relatively more exotic
(non-$q\bar{q}$) mesons, such as glueballs, $(q\bar{q}g)$ states,
etc. In other words the process $pp\ra p+X(2^{++})+p$ indeed acts
as a filter for separating out exotic mesons from normal mesons
\cite{CKIRK}.

The next observation is that mesons produced in the process $pp\ra
p+X+p$ by Pomeron-Pomeron fusion have larger transverse momenta,
$P_t$, where $P_t=|\vec{p}_{1t}+\vec{p}_{2t}|$. The reasons are
that (i)~the $J_z=0$ selection rule is absent at larger $p_{it}$,
(ii)~for a $1^{++}$ meson, the vertex contains a factor
$\epsilon^{\mu\nu\alpha\beta}p_{1t\mu}p_{2t\nu}$ and so prefers
larger $p_{it}$\footnote{By analogous arguments, the forward
production of the unnatural spin-parity states, $0^{-+}$ and
$2^{-+}$, should be strongly suppressed and also favours large
$p_{it}$. This does not depend on whether or not the mesons are
quarkonium states. Also, $1^{-+}$ production would tend to occur
at large $p_{it}$.}, and, finally, (iii)~heavier (exotic)
mesons\footnote{This, of course, is also valid for $0^{++}$
mesons.} tend to have larger $P_t$ due simply to kinematics.

Inspection of Fig.~3 shows that for small $p_{it}$ the cross
section decreases with increasing $\phi$, while for large $p_{it}$
it increases. In the latter domain the dominant contribution comes
from the back-to-back configuration.

All the qualitative features described above are indeed observed
in the available data \cite{WA102} for process (\ref{eq:a26}):

\begin{itemize}
\item[(i)] Pomeron-Pomeron fusion in double-diffractive meson production may act as a glueball filter: the final state is enriched by non-$q\bar{q}$ mesons, which have smaller $P_T$ and are produced mainly with tagged protons in the $\phi = 0$ configuration;
\item[(ii)] the normal $q\bar{q}$ light mesons have larger $P_T$ and their cross sections peak at $\phi=180^\circ\:$\footnote{The preference for double-diffractive $f_2$(1270)-meson production in the $\phi>90^\circ$ domain has been observed at higher energies at the ISR \cite{ISR}.};
\item[(iii)] the $2^{++}$ channel is produced mainly in the $J_z=0$ state \cite{WA2}.
\end{itemize}

These expectations can be confirmed by observing the
double-diffractive production of heavier quarkonia, like $\chi_c$
and $\chi_b$, at the Tevatron and at RHIC. The heavier mesons
sample smaller distances and so their production should be better
described by perturbative QCD. Of course, $\chi_c$ is probably
still not heavy enough, but nevertheless it would be interesting
to compare $\chi_c(2^{++})$ production with the enhanced $2^{++}$
glueball production rate.

\section{Conclusions}

It is well known that, in general, absorptive effects in inelastic
diffractive processes are much stronger than in the elastic
amplitude (see, for example, \cite{WK}). Such rescattering clearly
violates Regge factorization and leads to non-trivial correlations
between the transverse momenta $\vec{p}_{1t}$ and $\vec{p}_{2t}$
of the forward going protons in processes of the type $pp\ra
p+M+p$. Measuring the $p_{it}$ and the azimuthal angle $\phi$
distributions can provide an interesting possibility to probe the
opacity $\Omega(s,b_t)$ of the incoming proton and, moreover, to
test the dynamics of soft rescattering.

One of the best examples to study these effects is exclusive
high-$E_T$ dijet production, $pp\ra p+{\rm dijet}+p$, where the
cross section for the hard subprocess is large and well known.

Although questionable, the above perturbative formalism was
applied to central double-diffractive meson production at lower
energies, at which data exist. Surprisingly, the qualitative
features of these data were reproduced.

\section*{Acknowledgements}

We thank Frank Close, J. Lamsa, Risto Orava and Albert de Roeck
for interesting discussions. One of us (VAK) thanks the Leverhulme
Trust for a Fellowship. This work was partially supported by the
UK Particle Physics and Astronomy Research Council, by the Russian
Fund for Fundamental Research (grants 01-02-17095 and 00-15-96610)
and by the EU Framework TMR programme, contract FMRX-CT98-0194 (DG
12-MIHT).





\vfill

\newpage


\end{document}